\documentclass[pra,aps,epsf,twocolumn,nofootinbib,notitlepage,showpacs,10pt]{revtex4-1}
\usepackage{amsmath, verbatim,graphicx}
\usepackage{amsfonts}
\usepackage{color,soul}
\newcommand{\bra}[1]{\langle {#1} |}
\newcommand{\ket}[1]{| {#1} \rangle}
\newcommand{\expect}[1]{\langle {#1} \rangle}

\begin{document}

\title{Anti-bunching dynamics of plasmonically mediated entanglement generation}
\author{Eugene Dumitrescu$^{1,2,3}$ and Benjamin Lawrie$^{1,2,3}$}

\affiliation{$^1$ Quantum Computing Institute, Oak Ridge National Laboratory, Oak Ridge, TN 37831 \\
             $^2$ Quantum Information Science Group, Oak Ridge National Laboratory, Oak Ridge, TN 37831 \\
             $^3$ Bredesen Center for Interdisciplinary Research, University of Tennessee, Knoxville, TN 37996 }
\begin{abstract}
Dissipative entanglement generation protocols embrace environmental interactions in order to generate long-lived entangled states.
In this letter, we report on the anti-bunching dynamics for a pair of actively driven quantum emitters coupled to a shared dissipative plasmonic reservoir.
We find that anti-bunching is a universal signature for entangled states generated by dissipative means and examine its use as an entanglement diagnostic.
We discuss the experimental validation of plasmonically mediated entanglement generation by Hanbury Brown-Twiss interferometry with picosecond timing resolution
determined by an effective two-qubit Rabi frequency, and we analyze the robustness of entanglement generation with respect to perturbations in local detunings, couplings, and driving fields.
\end{abstract}

\maketitle

% ORNL
\footnotetext{This manuscript has been authored by UT-Battelle, LLC, under Contract No. DE-AC0500OR22725 with the U.S. Department of Energy.
The United States Government retains and the publisher, by accepting the article for publication, acknowledges that the United States
Government retains a non-exclusive, paid-up, irrevocable, world-wide license to publish or reproduce the published form of this manuscript,
or allow others to do so, for the United States Government purposes. The Department of Energy will provide public access to these results
of federally sponsored research in accordance with the DOE Public Access Plan.}

\section{Introduction}
Quantum decoherence results from interactions with unknown or uncontrollable environmental degrees of freedom.
This process, by which quantum information deteriorates due to environmental interactions, has been coined information leakage\cite{Nielsen}.
It follows that quantum information processing systems should be completely isolated from leaky environments.
However, such a task has proven to be quite difficult, and as a consequence, a variety of techniques have been developed to combat the effects of decoherence\cite{Terhal_RMP}.
With the fault tolerant threshold theorem \cite{Aharonov_98} providing a route to overcome decoherence,
quantum error correction protocols \cite{Shor, Calderbank, Laflamme_96, Steane, Gottesman_97, Kitaev_97, Terhal_RMP}, dynamical decoupling protocols \cite{Lorenza_DD},
and decoherence suppressing quantum control techniques \cite{Doherty_2000, Dong_Review} have all seen substantial progress.

% DDE
Dissipative driven entanglement (DDE) techniques provide a different and complimentary route to quantum state engineering \cite{Verstraete_Nat_Phys}.
In this paradigm, entanglement is stabilized\cite{Reiter_NJP} and computations are performed\cite{Kastoryano_PRL_13} by leveraging select dissipative pathways that are naively assumed to impede long term quantum coherence.
Early experimental DDE progress has been achieved in trapped ion \cite{Lin_Nature}, atomic ensemble \cite{Muschik_2011, Krauter_2001}, and superconducting \cite{Liu_2016} qubit platforms.

%Plasmons
Concurrently, a new quantum information processing platform based on the quantum theory of plasmons has rapidly matured in recent years \cite{Tame_Review, Ginzburg_Review}.
The first demonstration of plasmonically mediated entanglement \cite{Altewischer_Nature} stimulated developments in both discrete \cite{Tame_Review} and continuous \cite{Lawrie_2013, Holtferich_2016} plasmonic quantum variables.
More recently, squeezed states of light have enabled ultra-trace plasmonic sensing \cite{Fan_2015,Lawrie_2016},
while plasmonic mode volumes orders of magnitude below the diffraction limit have enabled Purcell factors exceeding $10^3$ in the weak coupling limit\cite{Akselrod_2014}
and vacuum Rabi splitting in the strong coupling limit \cite{Chikkaraddy_2016, santhosh2016vacuum}.
These plasmonic analogs to photonic cavity QED provide a framework for the development of nanoscale architectures with ultrafast coupling dynamics capable of operation at ambient temperatures.

% Motivation and Results
Despite substantial theoretical progress \cite{Tudela_2010, Cano_2011, Nori_2011, Tudela_2011, Lee_2013,
                                               Hummer_2013, Nerkararyan_2015, Ren_2015, Gangaraj_2015,
                                               Hou_15, Otten_2015, Otten_2016, Hakami_2016},
dissipative entanglement generation has yet to be observed in plasmonic platforms.
This is partially due to the technical difficulty of integrating plasmonic components with standard readout and control technologies.
It is therefore tremendously important to develop alternative yet simple entanglement metrics to develop the nascent field of plasmonic quantum information processing.
In this article we address this need by demonstrating how the second order temporal correlation function can be used as a signature of entanglement between a pair of qubits coupled to a common plasmonic environment.

We analytically and numerically treat the dynamics of the dual quantum dot - plasmon hybrid system, illustrated in Fig.~ 1, and analyze the photon anti-bunching as a function of steady state two qubit entanglement.
We also argue that an experimental demonstration is possible, despite the fast qubit timescales inherited from the plasmonic reservoir.
Importantly, we show that the two-qubit anti-bunching width can be classically tuned by controlling the external driving fields.
Specifically, reducing local driving amplitudes slows the effective two qubit Rabi frequency and extending the anti-bunching width to timescales as long as tens of picoseconds.
Picosecond timescales are currently experimentally accessible and further, they are orders of magnitude shorter than typical coherence times observed in anti-bunching measurements of single quantum emitters (point defects, quantum dots, etc.).
Anti-bunching lifetime measurements may therefore be used to distinguish between dissipative driven entangled systems and weakly interacting single emitters.

\begin{figure}
\includegraphics[width = \columnwidth]{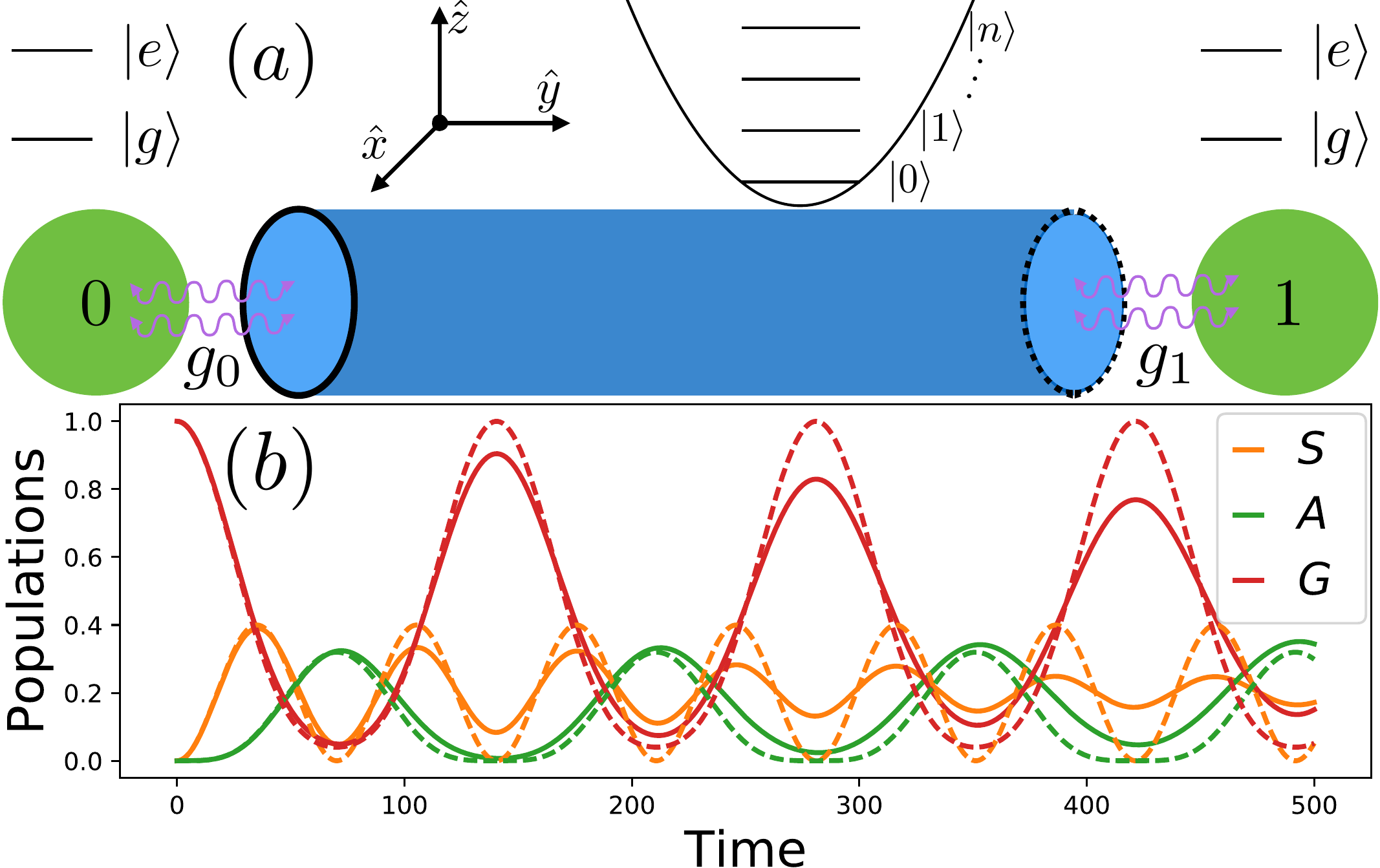}
\caption{
(a) Schematic diagram of the setup described by Eq.~\ref{eq:H_tot}.
(b) Solid (dashed) lines show the numerical (effective analytic) populations of the ground (G), symmetric (S), and antisymmetric (A) states for a system with Hamiltonian parameters (and their effective counterparts without dissipation):
 $\Delta_0 = - \Delta_1 = 0.02$, $g_0=g_1 = 0.02$, $\eta_0 = \eta_1 = 0.02$, and $\gamma_d = 2 \gamma_r = 10^{-8}$. Hamiltonian parameters in all figures are expressed as ratios with respect to the dominant energy scale set by $\gamma_a = 50 THz$.}
\label{fig:schematic}
\end{figure}

\section{Theory}
We consider a physical setup, illustrated in Fig.~\ref{fig:schematic}, consisting of a pair of qubits placed in close proximity to the near-field of a surface plasmon mode supported on a metallic nanowire.
Qubit-qubit interactions are thus mediated by a plasmonic boson reservoir.
The bare plasmon Hamiltonian is $H_{pls} = \int d \bm{r} \int_0^\infty d \omega \hbar \omega \hat{a}^\dagger(\bm{r},\omega) \hat{a}(\bm{r},\omega)$,
where $\hat{a}^{ }(\bm{r},\omega)$ and $ \hat{a}^{(\dagger)}(\bm{r},\omega)$ are the destruction and creation operators for elementary plasmonic excitations which satisfy bosonic commutation relations.
A resonant or near-resonant mode may be treated as an oscillator $H_{pls} = \hbar \omega_a \hat{a}^\dagger \hat{a}$ with a principal frequency $\omega_a$.
The qubit Hamiltonian reads $H_{i} = \hbar \omega_{i} \hat{\sigma}_i^{+} \hat{\sigma}_i^{-}$, where $i=0,1$ indexes the emitters which are modeled as two level systems with $\hat{\sigma}_i^{\pm}$ being the Pauli ladder operators, $\hat{\sigma}^{\pm} = \hat{\sigma}^x \pm i \hat{\sigma}^y = \ket{e}\bra{g} (\ket{g}\bra{e})$.
The qubits could be implemented by a variety of solid state platforms, for example, as semiconductor quantum dots\cite{Hanson_RMP, Stobbe_RMP}.
We do not restrict ourselves to a specific qubit platform, but note that our results are generally applicable given appropriate plasmonic mode matching, which may be tuned by adjusting the nanowire geometry \cite{Chang_2006, Dionne_2006, Benz_2016}.
Defining the plasmonic and TLS dipole operators as $\hat{d}_a = \hat{a}^\dagger + \hat{a}^{}$ and $\hat{d}_i = \hat{\sigma}^+_{i}+\hat{\sigma}^-_{i}$, the emitter-reservoir coupling is modeled by the interaction $H_{int} = \sum_{i} g_i \hat{d}_a \cdot \hat{d}_i$,
where $g_i \equiv (\mu_i \cdot E_i)/\hbar$ is the dipole interaction strength in which we have absorbed all physical constants, i.e. the emitter transition dipole moment $\mu_i$ and local plasmon electric field magnitude $E_i = \sqrt{\frac{\hbar \omega_a}{2 \epsilon_0 V}}$ where $V$ is the plasmon mode volume.
Plasmonic elements behave as lossy electromagnetic cavities in the both the weak and strong QED regimes \cite{Ginzburg_Review, Hummer_2013, Benz_2016,Chikkaraddy_2016, santhosh2016vacuum, Akselrod_2014}.
Finally, $H_{D} = - \sum_{i}( \eta_i e^{i \Omega_{i} t} \hat{\sigma}^{+}_i  + H.c.) - ( \eta_a e^{i \Omega_{a} t} \hat{a}^{\dagger}  + H.c.)$ models transitions being driven by external fields with amplitudes $\eta_{i(a)}$ and frequencies $\Omega_{i(a)}$.
Transforming to the co-rotating reference frame, with detunings $\Delta_{i(a)} = \hbar(\omega_{i(a)} - \Omega_{i(a)})$, and applying the rotating wave approximation, the total Hamiltonian becomes,
\begin{eqnarray}
\label{eq:H_tot}
H_{tot}  & = &  \sum_{i=0,1} \left[ \Delta_i \hat{\sigma}^+_i \hat{\sigma}^-_i - \eta_i \hat{d}_i  - g_i (\hat{\sigma}^+_i \hat{a} + \hat{\sigma}^-_i \hat{a}^\dagger) \right] \\ \nonumber
& + & \Delta_a \hat{a}^\dagger \hat{a}  - \eta_a \hat{d}_a.
\end{eqnarray}

Dissipation is modeled by treating the dynamics within the Lindblad master equation formalism,
\begin{equation}
\label{eq:ME}
\dot{\rho}= - i \left[H_{tot}  ,\rho \right] + \sum_{k} \gamma_k ( L_k \rho L_k^\dagger - \frac{1}{2}\{ L_k^\dagger L_k, \rho \})
\end{equation}
where we have take $\hbar = 1$ and the $L_k$ operators model various dissipative channels.
Specifically, we consider the following channels:
(i) plasmonic relaxation $L_a \equiv \hat{a}$ at a rate $\gamma_a \leq 50 \text{THz}$\cite{Rod_ACS},
(ii) emitter relaxation $L_{r,i} \equiv \hat{\sigma}_i^+$ at a rate  $\gamma_r = 2.5 \text{MHz}$, and
(iii) emitter dephasing $L_{d,i} \equiv \hat{\sigma}_i^z$ at a rate $\gamma_d = 5 \text{MHz}$.
Later we numerically solve Eq.~\ref{eq:ME} in full to validate anti-bunching and concurrence phenomenon in a wide range of parameter regimes.
However, we first derive an effective model to develop our intuition of the dynamics.  

\section{Two-qubit effective dynamics}
Let us now illustrate the dissipative flow dynamics for two qubits coupled through a common bosonic reservoir.
Both the effective model and our exact numerical calculations will identify the anti-symmetric singlet state $\ket{A} = \ket{eg} - \ket{ge}$ as a fixed point for the dynamical evolution in the parameter regimes highlighted below.
For our effective model, we work in the weak coupling regime defined by $ g_0, g_1,\eta_0, \eta_1 \ll \gamma_a$, where $\gamma_a$ is the rate for the relaxation channel taken with the relaxation timescale $\tau_a = 1/ (\pi*\gamma_a) \sim 6 fs$. 
%Physically this models the decay of plasmonic excitations due to coupling with vibrational and electronic modes. 
In this work we stay within this approximation so all energy and time scales are given as dimensionless ratios of $\gamma_a$. 
Still, the reservoir dynamics enables coherent communication channel between the distant qubits. 
As we now show, the system may be guided into the decoherence free subspace $\ket{A}$ by varying the qubits detunings and drivings in Eq.~\ref{eq:H_tot}, or equivalently, by driving the bosonic reservoir \cite{Hou_15}.

The effective qubit dynamics is found by the following adiabatic elimination procedure\cite{Hou_15}.
From Eqs.~\ref{eq:H_tot} and \ref{eq:ME} the Heisenberg equations of motion for the field operators are
\begin{eqnarray}
\label{eq:EOM_tot}
\dot{\sigma}^{z}_i &=&  i\left[ 2 g_i (\sigma^+_i a - \sigma^-_i a^\dagger) + 2 \eta_i (\sigma^+_i  - \sigma^-_i ) \right] \\ \nonumber
&-& \gamma_i (\mathbb{I} - \sigma_z)  + f^{z}_i \\ \nonumber
\dot{\sigma}^{-}_i &=&  - i \left[ \Delta_i  \sigma^-_i  + (g_i a + \eta_i) \sigma^z_i \right] - \gamma_i \sigma^-_i /2 + f^{-}_i  \\ \nonumber
\dot{a} &=& i \left[\eta_a - \Delta_a a +  g_0 \sigma^-_0 +  g_1 \sigma^-_1  \right] - \gamma_a a /2 + f_a  ,
\end{eqnarray}
with fluctuation operators $f^{z}_i, f^{-}_i, f_a$ representing higher order processes \cite{Hou_15}. 
Making the semi-classical approximation that expectation values for the fluctuation operators vanish, we decouple the expectation values of the qubits and the bosonic mode.
For slowly varying $\expect{a}$, valid in the case of weak coupling and drivings, we may set $\dot{a} = 0$ and substitute the resulting expression into the the first two lines of Eq.~\ref{eq:EOM_tot}.
This gives us the adiabatic Heisenberg equations of motion, which can in turn be viewed as arising from an effective two qubit Hamiltonian with non-local dissipation terms.
The effective Hamiltonian, $H_{qb} = \sum_{i} \left[ \tilde{\Delta}_i \hat{\sigma}^+_i \hat{\sigma}^-_i  -  \tilde{\eta}_i \hat{d}_i \right] - \tilde{g} (\hat{\sigma}^+_0 \hat{\sigma}^-_1  + \hat{\sigma}^+_1 \hat{\sigma}^-_0)$, is defined in terms of the following couplings: % (i.e., the qubits feel an effective driving field due to the presence of the plasmon)
(i) effective local detunings $\tilde{\Delta_i} = \Delta_i - g_i^2 \Delta_a/Z $,
(ii) effective driving fields  $\tilde{\eta_i} = \eta_i + g_i \Delta_a \eta_a/Z$, and
(iii) effective inter-qubit coupling $\tilde{g} = g_0 g_1 \Delta_a /Z$, where $Z = (\gamma_a/2)^2 + \Delta_a^2$.
The effective dissipations are described by $\sum_{ij=0,1} \tilde{\gamma}_{ij}/2 \left[ 2 \sigma^-_i \rho \sigma^+_j - \{ \sigma^+_j \sigma^-_i, \rho \} \right]$,
with single qubit relaxations occurring at a renormalized rate $\tilde{\gamma}_{ii} = \gamma_{i} + g_i^2 \gamma_a/Z$ and collective reservoir mediated relaxations occurring at the rate $\tilde{\gamma}_{ij} = g_0 g_1 \gamma_a/Z$.
Transforming to the Dicke basis, $\ket{E} = \ket{ee} \equiv \ket{s=1, m=1}$, $\ket{S} = \ket{eg} + \ket{ge}$, $\ket{A} = \ket{eg} - \ket{ge}$, $\ket{G} = \ket{gg}$, and defining the effective (anti)symmetric drivings $\eta_\pm = (\tilde{\eta}_0 \pm \tilde{\eta}_1)/\sqrt{2}$ and energies $\Delta_\pm = (\tilde{\Delta}_0 \pm \tilde{\Delta}_1)/2$.
Note that $\ket{A}$ is the part of the two-qubit space while $\ket{G}, \ket{S}, \ket{E}$ define the triplet basis vectors defined by angular momentum eigenvalues $m = -1, 0, 1$ respectively.

The Hamiltonian now reads
\begin{eqnarray}
\label{eq:H_D}
H_{D} & = &  \Delta_E \ket{E}\bra{E} + \Delta_S \ket{S}\bra{S} + \Delta_A \ket{A}\bra{A}  \\ \nonumber
& + & \Delta_- (\ket{A}\bra{S} + \ket{S}\bra{A}) \\ \nonumber
& - & \eta_- (\ket{S}\bra{G} + \ket{S}\bra{E} + H.c) \\ \nonumber
& - & \eta_+ (\ket{A}\bra{G} - \ket{A}\bra{E} + H.c). \\ \nonumber
\end{eqnarray}
where $ \Delta_E = 2 \Delta_+$, $\Delta_S = \Delta_+ + \tilde{g}$, and $\Delta_A= \Delta_+ - \tilde{g}$.

As a special case, we take anti-symmetric detunings and identical drivings, thus reducing the effective parameters to $\tilde{\Delta_i} = \Delta_i, \; \tilde{\eta_i} = \eta_i, \; \tilde{g} = 0, \; \tilde{\gamma}_{ii}  = \gamma_i + 4 g_i^2/\gamma_a,$ and $ \tilde{\gamma_{ij}} =   4 g_0 g_1  /\gamma_a$.
The resulting pure state populations, e.g. with $\rho(0) = \ket{G}\bra{G}$, oscillate with an effective two qubit Rabi frequency $\Omega = \sqrt{(\Delta/2)^2 + \eta^2}$ as
$\rho_E(t) = \left(\Delta ^2+\eta ^2 \cos (2 t \Omega )-\Omega ^2\right)^2/4 \Omega ^4, \; \rho_S(t) = \eta ^2 \sin ^2(2 t \Omega )/2 \Omega ^2, \; \rho_A(t) = 2 \Delta ^2 \eta ^2 \sin ^4(t \Omega )/\Omega ^4, \; \rho_G(t) = \left(\Delta ^2+\eta ^2 \cos (2 t \Omega )+\Omega ^2\right)^2/4 \Omega ^4$, and are illustrated by the dashed lines in Fig.~\ref{fig:schematic}.

% steady state |A><A|
The solid populations in Fig.~\ref{fig:schematic} are calculated by numerically solving Eq.~\ref{eq:ME} for the full system and can be understood as follows.
As discussed above, the excited state $\ket{E}$ relaxes to the singly excited state $\ket{eg(ge)}$ at a rate $\tilde{\gamma}_{00(11)}$.
However, in the Dicke basis, relaxation from the bi-excited state to the symmetric (antisymmetric) state occurs at the rate $\gamma_{S(A)} = \sum_i \tilde{\gamma}_{ii}/2 \pm \gamma_{ij}$.
Ignoring Hamiltonian dynamics for the moment, the populations are coupled as $\dot{\rho}_{SS} = (\rho_{EE} - \rho_{SS})  \gamma_{S}$ and $\dot{\rho}_{AA} = (\rho_{EE} - \rho_{AA})  \gamma_{A}$.
Thus, the solid lines in Fig.\ref{fig:schematic} and the eventual steady states, are driven by symmetric pumping described by Eq.~\ref{eq:H_D} augmented by super- and sub-radiant dissipation from the states $\ket{S},
$ and $\ket{A}$ respectively.
As discussed below, the entangled state $\ket{A}$ is a fixed point solution to these dynamics.

\begin{figure}[tb!]
\includegraphics[width = \columnwidth]{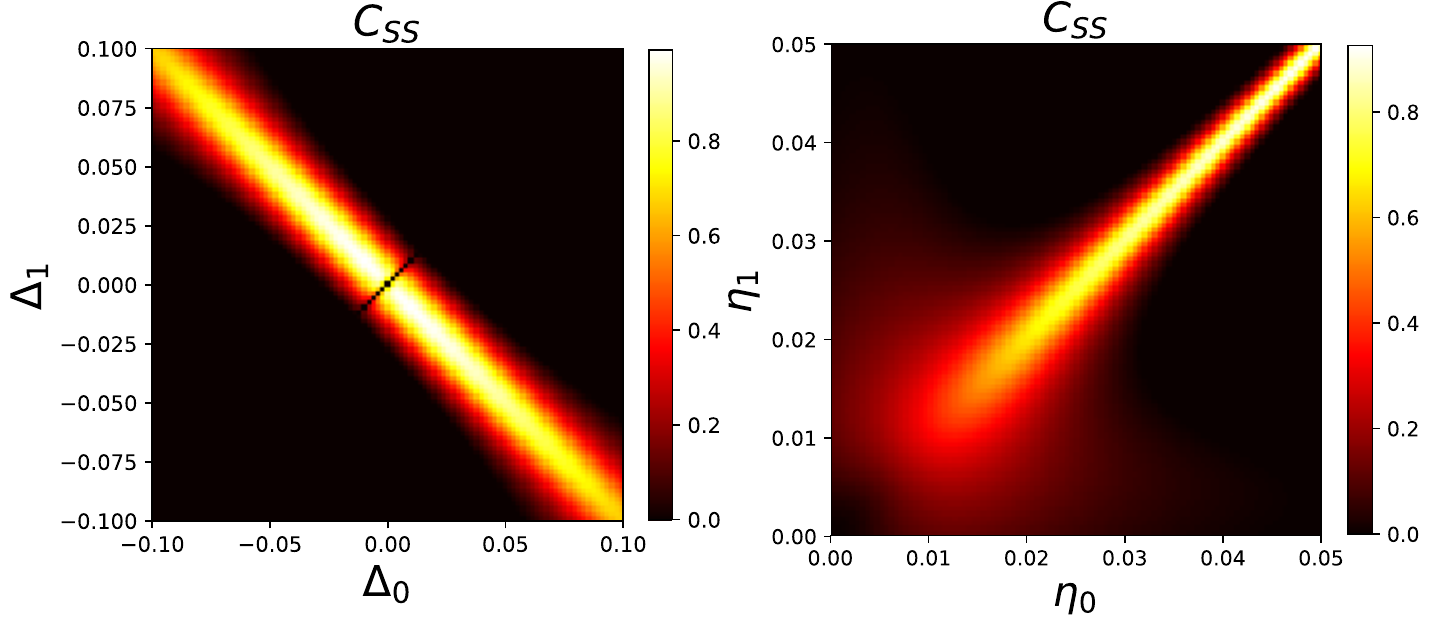}
\caption{Heat map of steady state concurrence $C_{ss}$ as a function of (left) qubit detunings $\Delta_{0,1}$ with equal qubit drivings $\eta_0=\eta_1 = 0.05$ and (right) qubit driving amplitudes $\eta_{0,1}$ with asymmetric qubit detunings $\Delta_0= - \Delta_1 = 0.01$.
Symmetric drivings and couplings with anti-symmetric detunings yields near unity concurrence steady states $\rho_{ss} \approx \ket{A}\bra{A}$.
Parameters common to both panels are coupling strengths $g_0 = g_1 = 0.05$ and plasmon detuning and driving $\Delta_a = \eta_a = 0$.}
\label{fig:concurrence}
\end{figure}

We now explore steady state characteristics as a function of the parameter space defined in Eq.~\ref{eq:H_tot}.
Fixed point entanglement is characterized by Wooter's concurrence $C$ \cite{Wooters_98}.
The concurrence of a two qubit state $\rho$ is $C(\rho) = \max\{0,\lambda_1 - \lambda_2 - \lambda_3 - \lambda_4\}$ where $\lambda_j$ are the sorted eigenvalues of $\rho \tilde{\rho}$, with the spin-flipped conjugate state $\tilde{\rho} = \sigma^y_1 \sigma^y_2 \rho^* \sigma^y_2 \sigma^y_1$.
$C$ ranges between $0$, for product states, and $1$, for maximally entangled states.
As indicated by our earlier discussion and illustrated in Fig.~\ref{fig:concurrence}, unit concurrence is readily achievable for systems with approximately equal couplings and driving fields as well as approximately anti-symmetric qubit detuning.
In all cases, the pair of qubits evolves to the anti-symmetric entangled state $\ket{A}$ \cite{Cano_2011, Hou_15}.

\begin{figure}[tb!]
\includegraphics[width =  \columnwidth]{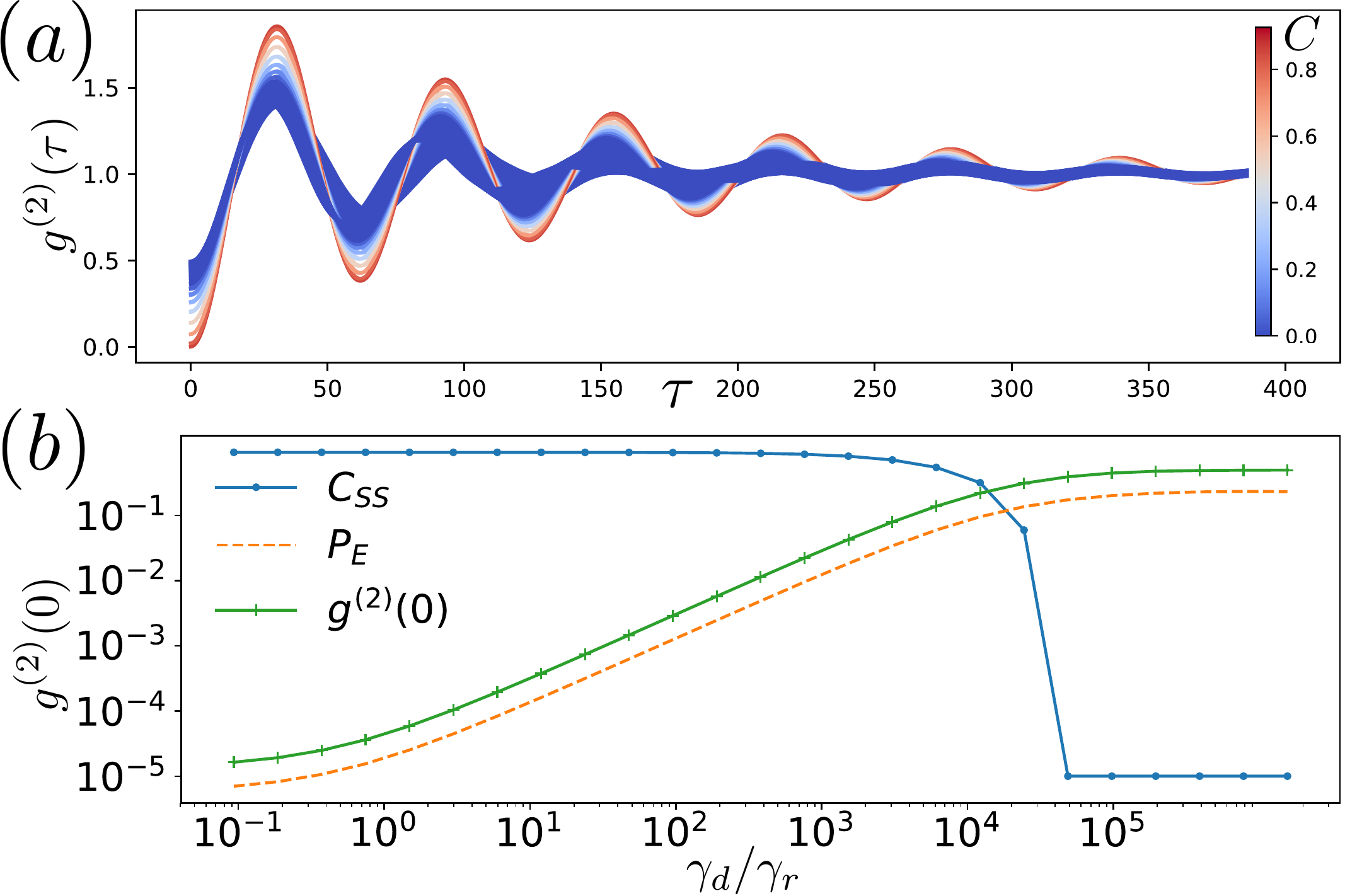}
\caption{
The steady state $g^{(2)}(\tau)$ correlation function as driving amplitudes $\eta_1 = 0.05$ and $\eta_0$ varies across $\left[0.04, 0.06\right]$.
The blue traces in panel (a) correspond to statistical mixtures of $\ket{gg}$ and $\ket{eg(ge)}$ depending on the circle endpoints.  
These states showing a $g^{(2)}(0)  = 0.5$ and correspond to general mixed states.
Over this range the concurrence evolves (see colorbar labeled by $C$) from $0$, where $g^{(2)}(0) = 1/2$, to $1$, where $g^{(2)}(0)=0$.
Other Hamiltonian parameters, as ratios of $\gamma_a$, are $\Delta_0 = -\Delta_1  = 0.02$, $g_0=g_1 = 0.05$, and $\eta_1 = 0.05$. $\tau$ is presented in units of $1/ (\pi*\gamma_a) \approx 6 fs$.}
\label{fig:g2}
\end{figure}

% g2 section
\section{$g^{(2)}(\tau)$ as an entanglement witness}
We now investigate the use of the second order temporal correlation function to quantify entanglement generation in our system.
Entanglement is typically validated by an ensemble of computational basis state measurements that are classically post-processed to either perform state tomography\cite{Poyatos_97, Chuang_97} or demonstrate a quantum inequality violation\cite{Bell, CHSH}.
Tomographic state readout has been successfully performed for dissipatively entangled trapped ions \cite{Lin_Nature} using specialized readout mechanisms.
However, for nascent plasmonic technologies, it is worthwhile to develop simple experimental signatures consistent with entangled states, without the complicated readout electronics needed to perform full state tomography.

In this context, anti-bunching in $g^{(2)}(\tau)$ of emitted light has been suggested as an alternative entanglement signature \cite{Cano_2011, Zheng_2013, Thakkar_2015}.
Below we confirm that the second order correlation function successfully discriminates between entangled, arising in the form $\ket{A}$, and unentangled steady states generated by our protocol.
Importantly, anti-bunching by itself is not conclusive evidence of entanglement between qubits with a shared dissipative pathway.  For instance, if both qubits were not well-coupled to the same plasmonic mode, steady-state entanglement would not be generated, but each qubit would exhibit anti-bunching on a timescale determined by the lifetime of the qubit.
 By considering the anti-bunching dynamics, it is possible to distinguish anti-bunching due to individual uncoupled emitters and anti-bunching due to dissipative, driven, entanglement between qubits coupled to a shared plasmonic reservoir.

The second order correlation function measures the degree to which a system is temporally correlated.
For stationary processes invariant under time translation, as is the case for steady states, the correlation function is defined as
\begin{equation}
\label{g2_tau}
\displaystyle g^{(2)}(\tau) = \frac{\langle a^\dagger (t) a^\dagger (t + \tau) a(t + \tau) a(t) \rangle}{\langle a^\dagger(t) a(t) \rangle^2}.
\end{equation}
In the context of quantum optics, $g^{(2)}(\tau)$ has the simple and intuitive interpretation of the normalized probability that two photons, whose emission times differ by $\tau$, are detected at a point in space.
A Hanbury Brown-Twiss (HBT) interferometer \cite{HBT} can be used to measure this quantity.
In a modern HBT interferometer, a 50/50 beamsplitter is used to send a light source to a pair of single photon counting detectors, and high speed electronics tag the arrival times of photons at each detector, with temporal resolution as fast as 1 picosecond.

In order to calculate $g^{(2)}(\tau)$, let us now combine continuous time evolution with a discrete quantum jump model.
Consider a steady state $\rho_{ss}$ of Eq.~\ref{eq:ME}, whose concurrence is plotted in Fig.~\ref{fig:concurrence}, which spontaneously emits a single photon from either qubit.
An emission event originating from the $i^\text{th}$ qubit corresponds mathematically to the application of a destruction operator $\sigma^{-}_i$ which projectively maps the post-emission state to $\rho_i(0) = \hat{\sigma}_i^{(-)}  \rho_{ss} \hat{\sigma}_i^{(+)} / \text{Tr} \left[ \hat{\sigma}_i^{(-)}  \rho_{ss} \hat{\sigma}_i^{(+)} \right]$.
Defining $\rho_i(\tau)$ as $\rho_i(0)$ evolved from $t = 0$ to $t = \tau$ according to Eq.~\ref{eq:ME}, the probability for a second emission from the $j$th qubit at time $\tau$ is then $\text{Tr} \left[ \hat{n}_j  \rho_i(\tau) \right]$, where $\hat{n}_j = \hat{\sigma}_j^{(+)}\hat{\sigma}_j^{(-)}$ is the qubit number operator.
Tracing over all emission configurations gives us the correlation function
\begin{equation}
\label{eq:g2_calc}
 g^{(2)}(\tau) = \sum_{ij} \text{Tr} \left[ \hat{n}_j  \rho_i(\tau) \right].
\end{equation}

\begin{figure}[tb!]
\includegraphics[width = \columnwidth]{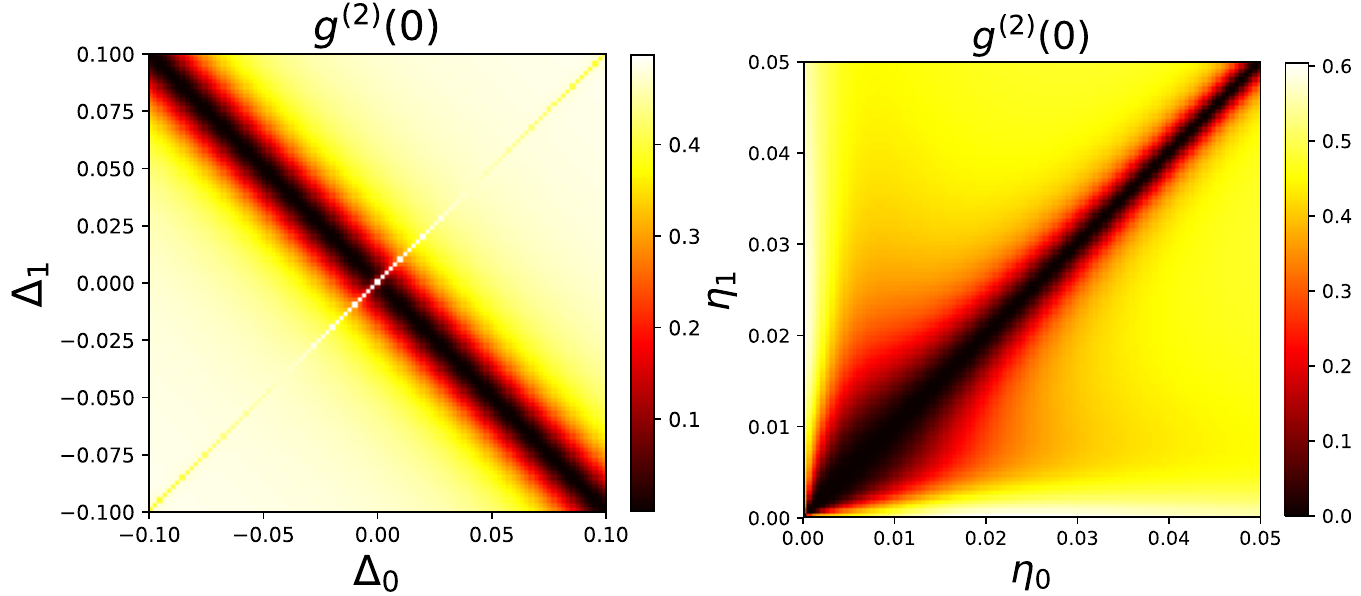}
\caption{Zero-delay correlations $g^{(2)}(0)$ as a function of (left) qubit detunings $\Delta_{0,1}$  and (right) qubit drive amplitudes $\eta_{0,1}$ using parameters reported in Fig.~\ref{fig:concurrence}.
Dark bands, which overlap strongly with the high concurrence regions, denote parameter regimes for which anti-bunching is present.}
\label{fig:anti-bunching}
\end{figure}

% g2(tau) figure/anti-bunching signature explanation
Fig.~\ref{fig:g2} panel (a) illustrates the behavior of $g^{(2)}(\tau)$ when the driving amplitude $\eta_0$ varies across $\left[0.04, 0.06\right]$.
Across this range, the concurrence varies from 0 to 1, and back to 0, as denoted by the color of the curves (also see Fig.~\ref{fig:concurrence}).
At unity concurrence we observe that $g^{(2)}(0) = 0 $, while for unentangled states, marked by vanishing concurrence, the zero-delay correlations saturate to $g^{(2)}(0) \sim 0.5$. 
Generally, this correlated anti-bunching signature appears for all dissipatively generated entangled steady states, as can be seen by comparing Figs.~\ref{fig:concurrence},\ref{fig:anti-bunching}.

It is interesting and necessary to study the effects of generic decoherence with respect to entanglement generation and anti-bunching. 
For example, we consider dephasing noise which is modeled by the presence of dephasing channels acting locally on each qubit.
In Fig.~\ref{fig:g2} panel (b) we vary the strength of the dephasing noise with respect to the relaxation rate and plot the steady state concurrence, zero-delay signal and population of the dually excited state $\ket{E} =  \ket{ee}$ 

% entanglement vs purity

We emphasize that the observation of an anti-bunching dip is not a general entanglement metric for arbitrary quantum states. 
However it is a universal feature shared by sub-radiance generated entangled steady states in our setup.
It is worth noting that anti-bunching is routinely observed in experiments involving single quantum emitters. 
The anti-bunching from single quantum emitters is rooted in the fact that after emitting a photon the emitter relaxes to its ground state and cannot source another photon without some time passing for the emitter to become excited again. 
For single quantum emitters, the anti-bunching dip width is proportional to the bare emitter decay rate ($\gamma_r$). 
Below we discuss how the width of the anti-bunching emanating from a pair of emitters, which are coupled by a common plasmonic reservoir, is many orders of magnitude smaller than a signal being sourced by a single quantum emitter.

Note that the robustness of the anti-bunching signal is rooted in the fact that, similar to the single emitter case, a single quantum is shared between two qubits in the form of the state $\ket{A}$.
Anti-bunching could also be caused by product states sharing a single quantum, e.g. $\ket{eg}$ or $\ket{ge}$, but these are unstable under the dynamics considered and the timescales would be quite different as already mentioned.
Further, $g^{(2)}(\tau)$ is unaffected by states including statistical mixtures of $\ket{G}$ (which don't contribute any emissions), while the bi-excited state $\ket{E}$ may generate two emission events with a small delay with high probability. These dual emissions destroy the anti-bunching signal.
Anti-bunching is therefore maximal in our setup for the only stable single excitation subspace: $\ket{A}$.

\begin{figure}[tb!]
\includegraphics[width = \columnwidth]{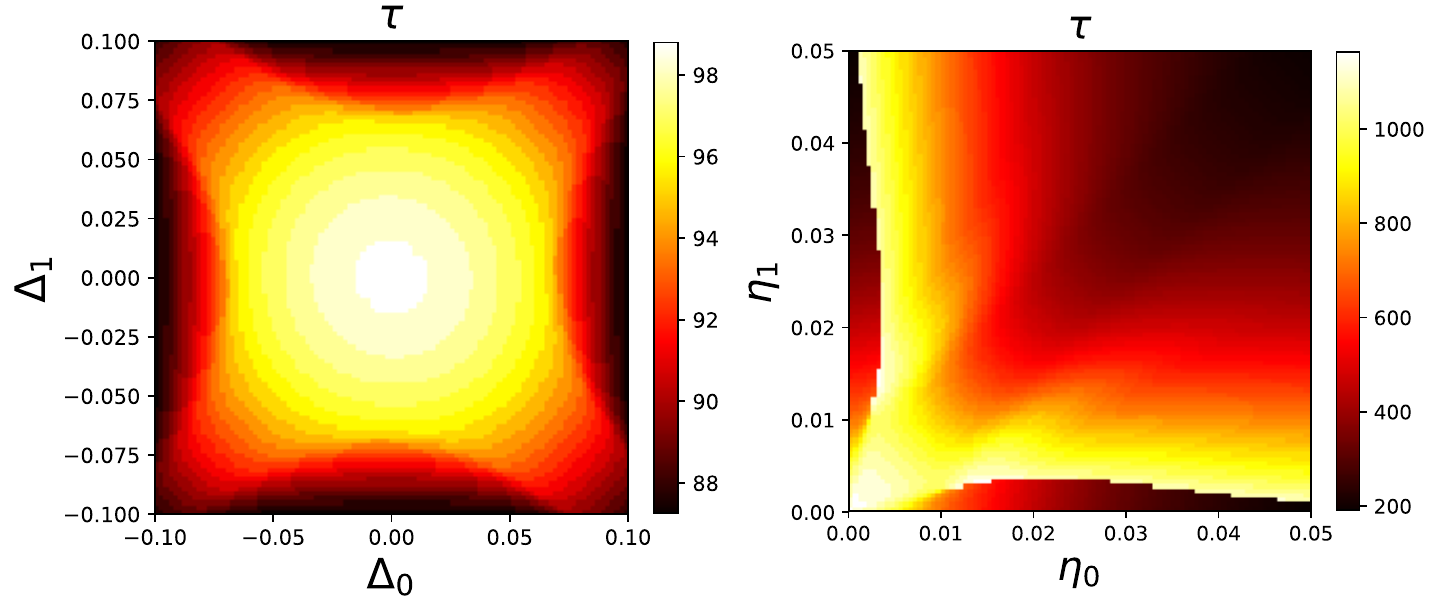}
\caption{Oscillation timescales for anti-bunching signal as a function of (left) qubit detunings $\Delta_{0,1}$  and (right) qubit drive amplitudes $\eta_{0,1}$.
Timescales reported are in units of $T = 1/ (\pi*\gamma_a) \approx 6 fs$ with remaining parameters as in Fig.~\ref{fig:concurrence}.
Entangled region around $\eta_0 = \eta_1 = 0.03 $ displays a characteristic bunching timescale $T_{AB} \sim 10 ps$.}
\label{fig:timescales}
\end{figure}

% Finite delay and anti-bunching timescales
While anti-bunching is an attractive entanglement signature due to its simplicity, its observation is non-trivial due to the fast time-scales inherited from the plasmonic reservoir.
After each radiative decay event, the two qubit system flows back to its steady state solution as described earlier.
Hence the width of the $g^{(2)}(0)$ anti-bunching dip is inversely proportional to the population oscillation Rabi frequency $\Omega$.
Tunable driving frequencies are therefore critical to observing anti-bunching on experimentally accessible timescales.
While nonlinear mixing with femtosecond laser sources could enable the detection of sub-picosecond dynamics in $g^{(2)}(\tau)$, conventional HBT interferometry is limited by the $1$ ps temporal resolution of state-of-the-art time tagging electronics.
To that end, we numerically calculate the oscillation timescales by Fourier transforming the $g^{(2)}(\tau)$ signal into the frequency domain and identifying the characteristic driving frequency, which fixes the anti-bunching timescale.
The timescales are provided in the Fig.~\ref{fig:timescales} color-maps, as a function of detuning and driving amplitudes, with normalized time in units of $T = 1/ (\pi*\gamma_a) \approx 6 fs$ reported in the color legend.
Anti-bunching timescales in the $\sim 10 ps$ range for entangled states are easily experimentally realizable, e.g. for small driving fields around $\eta_0 = \eta_1 \approx 0.03$. Notably, these timescales are much shorter than the lifetimes of typical qubits.  For instance NV centers in diamond have lifetimes of order 10-30 ns \cite{doherty2013nitrogen}.

%{\indent{\em Conclusion and discussion}}---
\section{Conclusion and discussion}
In this letter we have examined the entanglement characteristics of steady states generated by a pair of qubits subject to a dissipative plasmonic reservoir.
We have found that maximally entangled steady states are routinely achievable by appropriately tuning qubit detunings, couplings and driving frequencies.
Further, the entanglement was found to be robust against small perturbations in the tuning parameters, which need only be approximately symmetric (couplings, drivings) or anti-symmetric (detunings) in order to generate high concurrence states.

We have also examined entanglement detection by an anti-bunching signature in the second order correlation function that is routinely measured by means of a Hanbury Brown-Twiss interferometer.
By correlating this effect with the steady state concurrence, we have analyzed how the correlation function at zero-delay may serve as a robust entanglement signature for dissipating coupled qubit systems.
Importantly, we have also demonstrated that dynamics driven by weak fields allows the anti-bunching signature to persist on $\mathcal{O} (ps)$ timescales that pave the way to experimental detection using currently available experimental techniques.
This robustness against microscopic perturbations and unentangled fixed point states cements the $g^{(2)}(\tau)$ correlation function as a simple and practical measure of steady state entanglement generation.

{\indent{\em Acknowledgements.}}---
The authors thank F. Mohiyaddin for discussions and careful reading of the manuscript.
E. D. acknowledges support from the Intelligence Community Postdoctoral Research Program.
Research sponsored by the Intelligence Community Postdoctoral Research Fellowship and the Laboratory Directed Research and Development Program of Oak Ridge National Laboratory, managed by UT-Battelle, LLC, for the U.S. Department of Energy.
This manuscript has been authored by UT-Battelle, LLC, under Contract No. DE- AC0500OR22725 with the U.S. Department of Energy.

\end{document}